\documentclass{iopart}  
\usepackage{epsfig,subfigure}

\newcommand{\DIR}{.}

\begin{document}

\title{A realistic two-lane traffic model for highway traffic}

\author{W~Knospe\dag, L~Santen\ddag, A~Schadschneider\S~and 
M~Schreckenberg\dag}

\address{\dag\ Theoretische Physik FB 10,
                  Gerhard-Mercator-Universit\"at Duisburg,
                  Lotharstr. 1, D-47048 Duisburg, Germany}
\address{\ddag Fachrichtung Theoretische Physik, Universit\"at des
Saarlandes, Postfach 151150, 66041 Saarbr\"ucken, Germany}

\address{\S\ Institut f\"ur Theoretische Physik,
                 Universit\"at zu K\"oln,
                 Z\"ulpicher Str. 77, D-50937 K\"oln, Germany}


\begin{abstract}
A two-lane extension of
a recently proposed cellular automaton model
for traffic flow is discussed.
The analysis focuses on the reproduction of the  
lane usage inversion and the density dependence of the number of lane 
changes. It is shown that the single-lane dynamics can be extended to
the two-lane case without changing the basic properties of
the model which are known to be in good agreement with empirical
single-vehicle data. Therefore it is possible to
reproduce various empirically observed two-lane phenomena, like the
synchronization of the 
lanes,  without fine-tuning of the model parameters. 
\end{abstract}


\section{Introduction}

In~\cite{knospe:jpa} a cellular automaton model for traffic flow is
proposed that is based on the Nagel and Schreckenberg (NaSch) 
model~\cite{nagel:jphysique} and 
that is capable to reproduce all of the empirically observed
traffic states~\cite{kerner:tgf99,Kerner2001}, i.e., free flow, wide
moving jams and especially synchronized traffic (see \cite{CSS,Hrev} for
recent reviews on the modeling of traffic flow). 
Moreover, the accordance of the model with the empirics 
can be found even on a microscopic level of
description. In~\cite{knospe2001} this degree of realism is
underscored by considering a more realistic simulation
setup, i.e., a simulation of a two-lane highway segment with an onramp.
Empirical observations suggest~\cite{Kerner2001,Neubert} that lane
changes are responsible for the occurrence of strong correlations of
the velocity and flow measurements between neighboring lanes in the
synchronized states and in wide moving jams. However, it was
concluded~\cite{knospe2001} that lane changes do only select rather
than generate traffic states. 

Although the results of~\cite{knospe2001} do not depend on the special
choice of the lane change update,  the application
of the model in more sophisticated topologies should be provided 
with realistic lane changing rules.
 
Much progress has been made in the simulation of large traffic
networks like the inner-city of Duisburg~\cite{Esser97a}, the
Dallas/Forth-Worth area~\cite{Rickert97a} and 
the highway network of North-Rhine-Westfalia~\cite{babsim}.
These simulations were based on the NaSch model
and use a detailed representation of
the systems infrastructure like multi-lane traffic, on- and offramps
and highway intersections. 
However, a realistic microscopic description of the dynamics of each
individual vehicle in  
a network requires to take into account the peculiarities of the
complex road structures.
It is therefore necessary to formulate simple rules in the language
of cellular automata 
which mimic the behavior of the drivers as simple
as possible and yield reasonable results compared with empirical findings.
Unfortunately it turned out that the incorporation of 
two-lane traffic even in the NaSch model is a
formidable task and the formulation of realistic lane changing rules
is very difficult.

Only a few empirical results for two-lane traffic
exist that help to specify lane changing
rules~\cite{sparmann,chang,hall,brackstone}. 
The ability to change lanes should increase with the density,
shows a maximum in the vicinity of the flow maximum and then decreases
with increasing density~\cite{sparmann}. Nevertheless, lane changes are
still possible at large densities.
A special feature of German highway traffic is the empirically
observed lane usage inversion. Although there is a right lane 
preference, the distribution of the flow becomes asymmetric and the
flow is larger on the left than on the right lane 
\cite{sparmann,leutzbach}. In
contrast, in urban traffic or on highways without the right lane
constraints the traffic flow is evenly distributed on both
lanes~\cite{chang,hall,brackstone}. 

Several two-lane extensions~\cite{rickert:physicaa,nagel:physreve,wagner:physicaa,wagner:tgf96,nagatani:tgf96,nagatani:jpa}
of the NaSch model
were proposed which are able to reproduce the empirically observed lane 
usage inversion. However, they fail to
model a realistic density dependence of the  number of lane changes. 
Moreover, it was shown that some 
problems  of the cellular automaton approach exist if one 
introduces different kinds of vehicles, but these shortcomings can be
minimized by 
the consideration of anticipation effects~\cite{knospe:physicaa}.

Here, we extend the recently introduced traffic model~\cite{knospe:jpa} 
to two-lane traffic and answer the question which mechanism gives rise
to the empirically observed lane usage inversion. 
It is clarified that this extension can be made without changing the
basic realistic properties of the single-lane model.
Moreover, it is demonstrated that
lane changes do lead to strong correlations between neighboring lanes.

The simulations are (if not stated otherwise) performed on a highway
of length $50 000$ cells with periodic boundary conditions. In
addition to the local measurements of~\cite{knospe:jpa} traffic data
are also recorded by averaging over the whole highway segment.  This
allows, in particular, to observe lane changes which cannot be
measured by local detectors.  The averaging process, however, does not
distinguish between the various traffic states on different parts of
the highway~\cite{Neubert}.

In the next section we briefly summarize the definition of the
single-lane model and present a simple two-lane extension.
Because it is not possible to
generate a lane usage inversion by the introduction of different kinds of
cars, we present in section~\ref{section:asym} an asymmetric lane
changing rule set. 
In section~\ref{section:local} we show that the vehicle dynamics
of the single-lane model is not influenced by the
lane-changes, but the car dynamics
is determined by the single-lane update rules only. The advantages of
two-lane traffic in comparison to single-lane traffic become obvious
in section~\ref{section:comparison}. Finally, we conclude
with section~\ref{section:conclusion}.


\section{Symmetric model}
\label{section:sym}

Before we will present our lane changing rule set, we briefly
define the single-lane model (see~\cite{knospe:jpa} for a detailed
explanation). The model used here is based on the
Nagel-Schreckenberg cellular automaton model for traffic
flow~\cite{nagel:jphysique}. The road is divided into cells and a
vehicle has a length of five cells. The movement of
the vehicles is determined by a set of simple update rules and the
system update is performed in parallel.
The model of \cite{knospe:jpa} is based on the desire of the
drivers for smooth and comfortable driving whereas the dynamics
of the NaSch model only assures the avoidance of crashes.
This more sophisticated behavior is incorporated in a driving strategy 
which comprises three aspects:
(i) At large distances the movement of a vehicle is
determined by the maximum possible velocity. Vehicles at rest have a smaller
acceleration capability than moving vehicles. Additionally, the
velocity of the leading car is anticipated which allows for smaller
gaps and larger velocities.
(ii) At intermediate
distances vehicles anticipate last minute braking by means of brake
lights.
(iii) At small distances the drivers adjust their velocity such that
safe driving is possible.

\begin{table}
\begin{tabular}{p{1.5cm}p{4.1cm}|p{1.5cm}p{4.1cm}}\hline
variable &  & parameter &\\ \hline
$x$ & position & $gap_{\rm{safety}}$ & controls the effectiveness of
the anticipation\\ 
$v$ & velocity & $v_{\rm{max}}$ & maximum velocity\\
$v_{\rm{anti}}$ & anticipated velocity & $p_{b}$ &deceleration  probability\\
$d$ & distance-headway & $p_{0}$ &deceleration probability\\
$d^{\rm{eff}}$ & effective distance-head\-way & $p_{d}$ &deceleration probability\\
$b$ & status of the brake light & $t_{h}$ & time-headway\\
$p$ & deceleration probability & $t_{s}$ & safety time-headway\\
\hline
\end{tabular}
\caption{Summary of the variables  and of the  parameters of the model
used in the model definition.} 
\label{tab}
\end{table}

Introducing the randomization function
\[
p(v_{n}(t),b_{n+1}(t),t_h,t_{s}) = \left\{\begin{array} {llll}
      p_b & &{\rm if\ }b_{n+1}=1 {\rm \ and\ } t_{h} < t_{s}\\ 
      p_0 & &{\rm if\ } v_n=0 {\rm~and~not~} (b_{n+1}=1 {\rm \ and\ } t_{h} < t_{s})\\
      p_d & &{\rm in\ all\ other\ cases}.
\end{array} \right.
\]
and the effective distance
\[
d_n^{{\rm eff}} = d_n + \max(v_{{\rm anti}}-gap_{{\rm safety}},0),
\] 
where $d_n$ is the gap of vehicle $n$, 
$v_{{\rm anti}} = \min(d_{n+1},v_{n+1})$ the expected velocity of the 
preceding vehicle in the next time step and $gap_{{\rm safety}}$ controls
the effectiveness of the anticipation,
the update rules consist of the following 5 steps (see table~\ref{tab}
for a summary of the parameters and variables of the model):
\begin{enumerate}
\item[0.] Determination of the randomization parameter $p$:\\
$p = p(v_n(t),b_{n+1}(t),t_h,t_{s})$\\
$b_n(t+1) = 0$
\item[1.] Acceleration:\\
if $((b_{n+1}(t) = 0)$ and $(b_{n}(t) = 0))$ or
$(t_{h} \ge t_{s})$ then:\\ \hspace*{0.6cm}$v_n(t+1) =
\min(v_n(t)+1,v_{{\rm max}})$

\item[2.] Braking rule:\\
$v_n(t+1) = \min(d_n^{{\rm (eff)}},v_n(t))$ \\
if $(v_n(t+1) < v_{n}(t))$ then: \hspace*{0.2cm}
$b_n(t+1) = 1$ 

\item[3.] Randomization and braking:\\
if $(rand() < p)$ then:\\
\hspace*{0.2cm} $v_{n}(t+1)
=\max(v_n(t+1)-1,0)$\\ 
\hspace*{0.3cm}if $(p = p_{b})$ then: $b_{n}(t+1) = 1$

\item[4.] Car motion:\\
$x_{n}(t+1) =x_{n}(t)+v_n(t+1)$\\

\end{enumerate}

Step $0$ determines the value of the randomization parameter $p$ which
depends on the status of the ``brake light'' $b_n$ (on/off) and the
velocity $v_{n}$ of the $n$-th car (Cars are numbered in the driving 
direction, i.e., vehicle $n+1$ precedes vehicle $n$. In case of two-lane
traffic this numbering is applied to each single lane
separately. Thus, cars are relabeled after lane changes.). 
In the next three steps the velocity $v_n$ of vehicle $n$ 
is calculated. A car accelerates if its own brake light and that of 
the leading car is switched off. If the time-headway $t_h$ to the next car is
large enough the car tries to approach the desired velocity in any
case. The braking rule ensures safe driving -- the velocity is adapted
to the effective distance $d_n^{{\rm (eff)}}$ to the leading vehicle. 
The effective distance is used instead of the real spatial 
distance $d_n$ to the leading vehicle. If the velocity is
reduced the brake light is switched on. The next step introduces the
stochastic element into the model, i.e., the velocity is reduced with 
probability $p(v_n(t),b_{n+1},t_h,t_{s})$ as determined in step 0. Please 
note that the implementation also allows for the propagation of the brake
lights.  In the final step the positions of the vehicles are
updated and the status of the brake light is reseted. The two times
$t_h = d_n/v_{n}(t)$ and $t_s = 
\min(v_{n}(t),h)$, where $h$ determines the range of interaction with the
brake light, are introduced to compare the time $t_h$ needed to reach
the position of the leading vehicle with a velocity-dependent
(temporal) interaction horizon $t_s$. 
Realistic behavior is observed for the model
parameters $v_{{\rm max}} = 20$, $p_d = 0.1$, $p_b = 0.94$, 
$p_0 = 0.5$, $h = 6$, $gap_{{\rm safety}} = 7$ which have been
used in the simulations. The cell length is $1.5~m$, a car has a
length of $5$ cells~\cite{knospe:jpa}. A time-step corresponds to $1$
s in reality.
Note that no fine-tuning of the parameters is necessary and the model
is quite robust against changes of the parameter values.

In order to extend the model to two-lane traffic one has to introduce
lane-changing rules. 
A lane changing rule set should (i) consist of simple, local
rules, (ii) be robust regarding slow vehicles, that is the flow should
not be dominated by the 
introduction of a small fraction of slow vehicles, (iii)
reproduce empirical 
lane changing curves, (iv) show the empirically observed lane usage
inversion (in the asymmetric case) and (v) should not change the 
dynamical behavior of the single-lane system.

In general, lane changing rules can be symmetric or asymmetric with
respect to the lanes or to the cars.
While symmetric rules treat both lanes equally (see, e.g., urban
traffic or highway traffic in the USA), asymmetric rule sets
especially have to be applied for the simulation of German highways,
where lane changes are dominated by a right lane preference and a
right lane overtaking ban. Moreover, an asymmetry between, e.g., cars
and trucks is provided on a two-lane highway since trucks are not
allowed to change to the left lane.

In principle, all lane changing rule sets of cellular automaton models
for traffic flow are formulated
analogously~\cite{nagel:physreve,CSS}.
First, a vehicle needs an incentive to change a lane. 
Second, a lane change is only possible if some
safety constraints are fulfilled.
Asymmetry is introduced if one applies different criteria for the
change from left to right and right to left.

A system update is performed in two sub-steps: In the first step the
cars change the lanes according to the lane changing rules and do not
move. In the second step, the cars move according to the calculated 
velocity. Both sub-steps are performed in parallel for all vehicles.

As a first step towards a realistic two-lane model for highway traffic
we studied symmetric lane changing rules with identical rule sets for 
the change from the left to the right and the right to the left 
lane. The lane changing strategy comprises two aspects:
First, the interaction of a vehicle with its predecessor on its lane
should be minimized for safety or just for comfort reasons. Thus, a vehicle
can optimize its travel time by driving as fast as possible due to an
optimal gap usage.
In order to keep the model as simple as possible we restrict 
the lane interaction to vehicles which have to brake
in the next time step due to an insufficient gap in front.
Vehicles which had to brake (i.e., their brake light is activated) are
not allowed to change the lane.
Second, safety reasons or constraints by law require
the minimization of the
interaction of a changing vehicle with its predecessor as well as
with its successor on the destination lane.
Thus, a vehicle is allowed to change the lane only if the gap between
successor and predecessor on the destination lane is sufficient. In
order to increase the efficiency of the lane changes, the movement
of the preceding vehicle on the destination lane is anticipated.

The lane changing rules then are as follows:

\begin{enumerate}
\item Incentive criterion:

\[ (b_n = 0)~and~(v(t) > d) \]

\item Safety criterion:

\[ (d^{{\rm (eff)}}_{{\rm pred}} \ge v(t))~and~(d_{{\rm succ}} 
\ge v_{{\rm succ}}) \]

\end{enumerate}

\begin{figure}[hbt]
\begin{center}
   \subfigure{\epsfig{file=\DIR/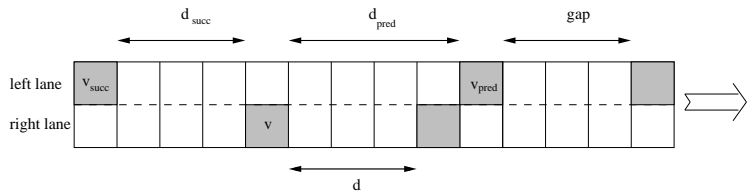, width=0.75\linewidth} }
\caption{Sketch of a road segment and illustration of the quantities
relevant for the lane changing rules. The hatched cells are occupied 
by a vehicle.}
\label{symmodel}
\end{center}
\end{figure}

$v(t)$ and $d$ are the velocity and the gap of the vehicle, 
$d_{{\rm succ}}$ and $v_{{\rm succ}}$ are the gap to the succeeding 
vehicle on the destination lane and its velocity, respectively.
$d_{{\rm pred}}$ denotes the gap and $d^{{\rm (eff)}}_{{\rm pred}}=
d_{{\rm pred}}+\max(v^{({\rm anti})}-gap_{{\rm safety}}, 0)$ denotes 
the effective gap to the preceding vehicle on the
destination lane (where $v^{({\rm anti})} = \min(gap,v_{{\rm pred}})$ 
is the expected velocity of the leading vehicle on the destination lane
in the next time step and $gap$ and $v_{{\rm pred}}$ is the gap and the
velocity of the predecessor on the destination lane) (see
Fig.~\ref{symmodel}). The
effectiveness of the anticipation is  
controlled by the parameter $gap_{{\rm safety}}$.
Accidents are avoided only if the constraint 
$gap_{{\rm safety}}\geq 1$ is fulfilled. 
The same value for $gap_{{\rm safety}}$ as in the single-lane model is 
applied.

\begin{figure}[hbt]
\begin{center}
\subfigure{\epsfig{file=\DIR/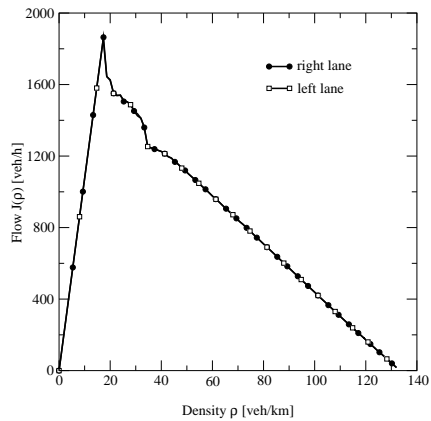, width=0.52\linewidth}}
\caption{Fundamental diagram of the individual lanes in the symmetric model.}
\label{symfundi}
\end{center}
\end{figure}

\begin{figure}[hbt]
\begin{center}
 \subfigure{\epsfig{file=\DIR/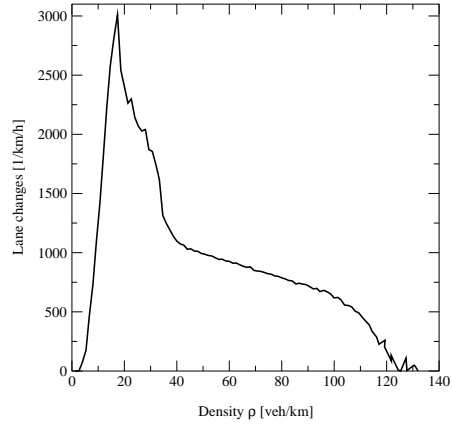, width=0.52\linewidth} }
\caption{Number of lane changes per lane in the symmetric model.}  
\label{symlanechanges}
\end{center}
\end{figure}

The symmetry of the lane changing update rules is reflected in the
fundamental diagram\footnote{Note, that the kink in the fundamental 
diagram is an artifact of the global measurement process: since it is 
not distinguished between the various traffic states on different 
highway sections, one simply averages over free flow and synchronized 
states.} (Fig.~\ref{symfundi}): both lanes show the same
flow density relationship, especially they have the same maximum
flow $J_{{\rm max}}$ at the same density $\rho_{{\rm max}}$.
Nevertheless, there is a strong exchange of vehicles between the lanes.
Although the empirical
lane change curve~\cite{sparmann} was taken from measurements on
German highways where asymmetric rules have to be applied, it is quite well 
reproduced on a qualitative level 
by simulations of the symmetric model while on a quantitative level the
number of lane changes is too large (Fig.~\ref{symlanechanges}). 
A detailed analysis reveals that at small
densities about $50\%$ of the lane changes are ping-pong
changes~\cite{rickert:physicaa}. Since anticipation is only applied in
step $2$ of the lane changing update, once a vehicle with an effective
gap changed the lane the condition $v > d$ of step $1$ is always
fulfilled. As a consequence, the vehicle changes the lane again.

\begin{figure}[h]
\begin{center}
 \subfigure{\epsfig{file=\DIR/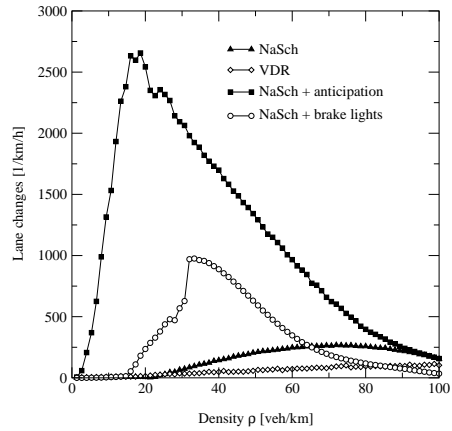, width=0.52\linewidth} }
\caption{Lane changes of successive extensions
of the symmetric two-lane NaSch model.}
\label{symlanechangesucc1}
\end{center}
\end{figure}

If we drop the single-lane model extensions (slow-to-start rule,
anticipation, brake lights) successively, the lane
change curve changes drastically (Fig.~\ref{symlanechangesucc1}).
The original NaSch model as well as the NaSch model with 
velocity-dependent randomization where a slow-to-start rule is applied
(VDR-model~\cite{barlovic:eurphys}) show a maximum of 
the lane change number at large densities, but not at
$\rho_{{\rm max}}$~\cite{sparmann}. 
Additionally, the number of lane changes is reduced considerably
compared to the full model (Fig.~\ref{symlanechanges}).
However, the introduction of brake lights or anticipation shifts
the maximum of the curve to $\rho_{{\rm max}}$.
On one hand, anticipation increases the effectiveness of the lane
changes because the gap acceptance is increased. Moreover, since
platoons of vehicles with a gap smaller than the velocity
exist, the first lane change rule is often fulfilled leading to an
artificial enhancement of the lane changing probability.
On the other hand, brake lights visualize breakdowns of the downstream
traffic timely, so that vehicles are able to swerve to the other lane
in order to avoid abrupt brakings.
Thus, from a microscopic point of view only the introduction of brake 
lights is responsible for the realistic lane changing behavior.

The symmetry of the lane changing rules is responsible for the even
distribution of the density on both lanes. 
In contrast, lane usage inversion was observed on German highways
where the left lane is designated as the passing lane and the lane
changing behavior is determined by the right lane preference and the
right lane overtaking ban.

In order to implement an artificial asymmetry we introduced
disorder by considering different types of vehicles like cars and
trucks.
Unfortunately, the introduction of disorder in cellular automaton
models for two-lane traffic has
some shortcomings: it is possible that two slow vehicles driving side
by side on different lanes can form a plug which blocks the succeeding
traffic~\cite{knospe:physicaa,chowdhury:physicaa}. These plugs are very 
stable in the free flow regime and their 
dissolution is determined by velocity fluctuations.
As a result, the lifetime of the plugs is very large and the flow is
dominated by the slowest vehicle. In the NaSch model without
anticipation, even one slow car can dominate the flow. 
Anticipation reduces the formation of plugs
considerably~\cite{knospe:physicaa}, but they still seem to be
slightly  overestimated by the model.
To avoid the formation of plugs all trucks are therefore initialized on
the right lane only, and are not allowed to change the lane.

\begin{figure}[hbt]
\begin{center}
 \subfigure{\epsfig{file=\DIR/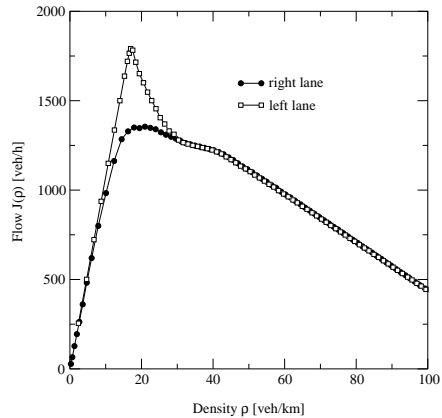, width=0.52\linewidth} }
\caption{Fundamental diagram of the individual lanes of the symmetric model
with $10\%$ trucks. Note, 
that the trucks are not allowed to change to the left lane.} 
\label{sym:inhomo:fundi}
\end{center}
\end{figure}

\begin{figure}[hbt]
\begin{center}
\subfigure{\epsfig{file=\DIR/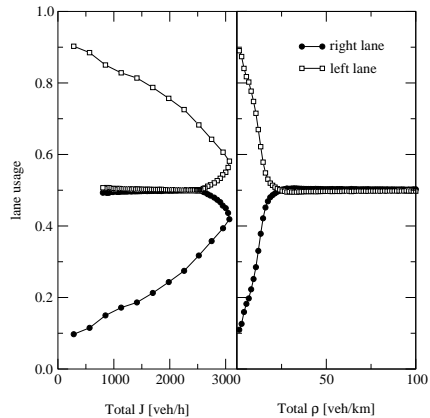,width=0.52\linewidth} 
} 
\caption{Lane usage of the symmetric model with $10\%$ trucks 
as a function of the flow (left) and the density (right).
Note, that the trucks are not allowed to change to the left lane.}
\label{sym:inhomo:laneusage:flow}
\end{center}
\end{figure}

Suppose that $10~\%$ of the vehicles are trucks with 
$v_{{\rm max}}^{{\rm trucks}} = 15 = 81~km/h$.
As a result of the asymmetric vehicle distribution the flow on the
right lane is dominated by trucks, while the flow on the left lane is
comparable to the flow of the homogeneous model (Fig.~\ref{sym:inhomo:fundi}).
Obviously, the introduction of trucks does not allow the reproduction
of the lane usage inversion (Fig.~\ref{sym:inhomo:laneusage:flow}): 
The flow on the left lane is larger than on the right lane for 
{\em all} densities since fast vehicles avoid to drive on the right lane.

However, since one value of the total flow can be related to either
free flow or congested traffic, the lane-usage is not a unique function of the
total flow. 
In contrast, the lane usage depending on the density gives 
unique results in the whole density regime.
Therefore, the lane usage curve described by the density is used in the
further analysis since, as an advantage, density can be
calculated explicitly in the simulations. 
Moreover, a comparison of both methods shows that the lane 
usage inversion point has the same flow-density value.

Due to the trucks on the right lane, at small densities 
nearly all fast cars are on the left lane. With increasing density the
velocity differences between the cars and the trucks decrease so that
both lanes are evenly occupied (Fig.~\ref{sym:inhomo:laneusage:flow}).
In contrast to the empirically observed lane usage inversion, neither
most of the cars are on the right lane at small densities, nor most of
the cars are on the left lane at large densities.
In order to obtain the lane usage inversion, it is therefore necessary
to incorporate an asymmetry into the lane changing
rules~\footnote{Note, that the lane usage inversion can be traced back
to an asymmetric distribution of the density on the lanes but not to a
negative velocity gradient from the left to the right lane that leads
to larger flows on the left lane. Different
speed limits for the lanes increase the inversion but are not its
reason, since lane usage inversion can also be observed on highways
with speed limit.}.

On German highways two mechanisms lead to an asymmetry between the
lanes:
\begin{enumerate}
\item[(i)] The {\em right lane preference} is enforced by the legal 
regulation to use the right lane as often as possible.
\item[(ii)] The {\em right lane overtaking ban} prohibits a car driving 
on the right lane to overtake a car which is driving on the left lane.
\end{enumerate}

In order to clarify which of these two mechanisms is responsible for
the lane usage inversion, we
first focus on the right lane overtaking ban.
Therefore, lane changes to the left lane are forced in the sense that  
a car on the right lane has to change to the left lane in case of a
velocity larger than its gap to the predecessor on the destination lane.
In addition to that, the safety criterion is weakened, so that cars even
with $v > d^{{\rm (eff)}}_{{\rm pred}}$ are allowed to change the
lane.
However, the overtaking ban is not formulated in a strict
sense since vehicles are allowed to overtake a car driving on the left
lane if a lane change is not possible.
As a result, most of the cars are on the left lane for all
densities. With increasing 
density this difference of the lane usage decreases, but
nevertheless it is not possible to observe a lane usage inversion.

Second, we focus on the right lane preference and force changes
from left to right. Obviously, the right
lane preference alone is not sufficient for the correct description of
the lane usage inversion, because cars now are predominantly on the
right rather than on the left lane.

To summarize, neither it is possible to get a lane usage inversion by
implementing disorder in the symmetric model, like different types of
cars, nor by introducing either the right lane 
preference or the right lane overtaking ban alone.
One therefore has to introduce the right lane preference {\em and}
the right lane overtaking ban simultaneously.

\section{Asymmetric model}
\label{section:asym}

The straightforward implementation of the
right lane overtaking ban of the preceding section leads
to large lane changing frequencies. 
Furthermore, a stricter formulation
of the right lane overtaking ban may lead to difficulties with
regard to velocity anticipation or to the update procedure.
For the sake of simplicity, however, 
the consequences of a right lane overtaking ban rather than the
basic mechanisms are modeled. Vehicles are still allowed to overtake
their predecessor on the left lane but the left lane should
be preferred. Therefore,
we 
reduced the ability to change from the left to
the right lane~\cite{nagel:physreve,wagner:physicaa} so that 
vehicles 
on the left lane change back to the right lane only if there is
a sufficient gap on both, the right and the left lane. 
The lane change rules are then as follows:\\

\begin{tabular}{p{3.5cm} p{7cm}}
& {\bf right $\rightarrow$ left}\\
& \\
Incentive criterion: & $ (b_n = 0)~and$ \\
& $(v(t) > d)$\\
& \\
Safety criterion: & $ (d^{{\rm (eff)}}_{{\rm pred}} \ge v(t))~and~(d_{{\rm succ}}
\ge v_{{\rm succ}}) $\\
& \\
\end{tabular}

\begin{tabular}{p{3.5cm} p{7cm}}
& {\bf left $\rightarrow$ right}\\
& \\
Incentive criterion:& $ (b_n = 0)~and$ \\
& $(t_{{\rm pred}}^{h} > 3.0)~and~((t^{h} > 6.0)~or~(v > gap)) $\\
& \\
Safety criterion: & $ (d^{{\rm (eff)}}_{{\rm pred}} \ge v(t))~and~(d_{{\rm succ}} \ge
v_{{\rm succ}}) $ \\ 
& \\
\end{tabular}

The two times $t^{h} = d/v$  and $t_{{\rm pred}}^{h} = d_{{\rm pred}}/v$ 
give the time a vehicle 
needs to reach the position of its predecessor and its predecessor on
the destination lane.
Since the time-headways take the velocity of the vehicles into account,
slow vehicles are allowed to change the lane even at
small distances~\footnote{Instead of comparing time-headways it is 
possible to use the vehicles gap reduced by a constant offset in the 
incentive criterion (e.g, $v < gap - offset$) in order to retard the 
change from the left to the right lane. Unfortunately, this has major 
drawbacks since for larger densities this condition is never be 
fulfilled~\cite{wagner:physicaa}.}.

\begin{figure}[hbt]
\begin{center}
\subfigure{\epsfig{file=\DIR/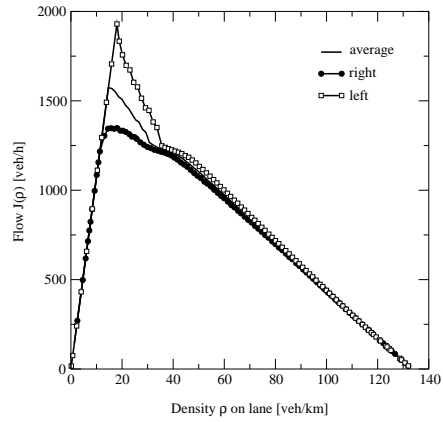,width=0.52\linewidth} } 
\caption{Fundamental diagram of the individual lanes in the asymmetric model.}
\label{asym:fundi}
\end{center}
\end{figure}

\begin{figure}[hbt]
\begin{center}
\subfigure{\epsfig{file=\DIR/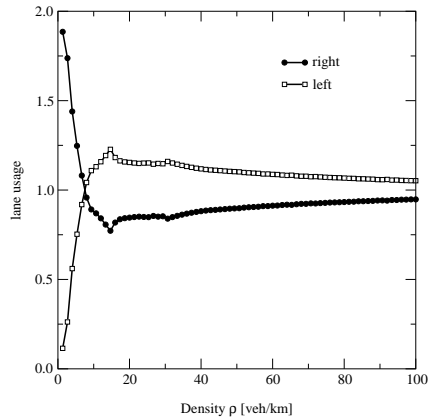,width=0.52\linewidth} } 
\caption{Lane usage in the asymmetric model.}
\label{asym:laneusage}
\end{center}
\end{figure}

With the two parameters $t_{{\rm pred}}^{h}$ and $t^{h}$ it is
possible to trigger the lane usage inversion.
If the movement of a vehicle is undisturbed (at small densities) it
changes the lane if  
its time-headway to the preceding vehicle on its own and on the
destination lane is larger than $6~s$ and
$3~s$, respectively.
Otherwise, at large densities the condition $v > d$ ensures
an incentive to change to the right lane.

Figure~\ref{asym:fundi} illustrates the asymmetry of the lane changing
rules with the fundamental diagram: at small densities, the 
flow on the left lane is larger than on the right lane. As a
consequence, the flow breakdown occurs first on the left lane.
Once 
the maximum flow on the left lane is reached more and more cars
swerve on the right lane until the flow is distributed evenly on both
lanes. Therefore, 
the system breakdown 
is mainly triggered by a single-lane breakdown on the left
lane~\cite{nagel:physreve}.  
However, the mean velocity on the right lane starts to decrease before
the breakdown on the left lane occurs, so that the velocity is always
larger on the left rather than on the right lane.
Thus, the system separates into a fast and a slow lane.

The two competing mechanisms of the lane changing rules, namely the
right lane overtaking ban and the right lane preference, lead to a lane
usage inversion (Fig.~\ref{asym:laneusage}).
At small densities the right lane preference dominates so that most of
the vehicles are driving on the right lane. However, as a consequence
of the right lane overtaking ban, with increasing
density more vehicles avoid driving on the right lane.
The position of the inversion point is mainly controlled by
$t^{h}$ while the extent of the lane usage inversion is controlled by
$t_{{\rm pred}}^{h}$. 
With increasing total density the lane usage inversion decreases
(Fig.~\ref{asym:laneusage}). 

\begin{figure}[hbt]
\begin{center}
\subfigure{\epsfig{file=\DIR/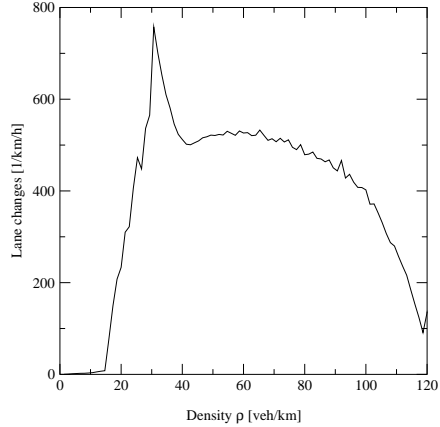,width=0.52\linewidth} } 
\caption{Number of lane changes in the asymmetric model.}
\label{asym:lanechangenumber}
\end{center}
\end{figure}

Due to the retarded lane change from left to right the number of lane
changes is decreased significantly compared to the symmetric model. 
Like in the symmetric model, the maximum of lane changes is reached in
the vicinity of the maximum flow. Further increasing the density the
lane change number drops to a plateau value until lane changes are no
longer possible at very large densities
(Fig.~\ref{asym:lanechangenumber}).
In contrast to the symmetric model, the ping-pong changes are strongly
suppressed due to the asymmetric rule set.

\begin{figure}[hbt]
\begin{center}
\subfigure{\epsfig{file=\DIR/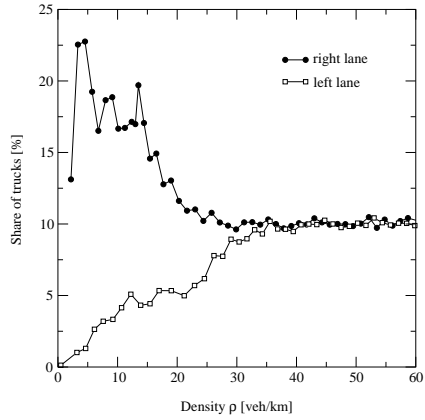,width=0.52\linewidth} } 
\caption{Distribution of the trucks on both lanes in the inhomogeneous
asymmetric model. Note, that trucks are allowed to change
the lanes.}
\label{asym:inhomo:sharenoban}
\end{center}
\end{figure}

Next, we consider an inhomogeneous
model with $10 \%$ trucks. In contrast to the symmetric model trucks
are now allowed to change from the right to the left lane.
In Fig.~\ref{asym:inhomo:sharenoban} the distribution of the trucks on
both lanes is depicted.
Although at small densities considerably more trucks are on the right
than on the left 
lane, the trucks do not  condensate completely on
the right lane. Therefore, the flow is dominated by
trucks on both lanes even for small densities and a small
fractions of trucks.
As a result of the lane usage inversion and a
small number of trucks on the left lane, vehicles can go faster on the
right rather than on the left lane.
However, at densities larger than
$\rho_{{\rm max}}$ the maximum velocity of the vehicles are smaller than the
mean velocity so that the trucks are evenly distributed on both lanes.

\begin{figure}[h]
\begin{center}
\subfigure{\epsfig{file=\DIR/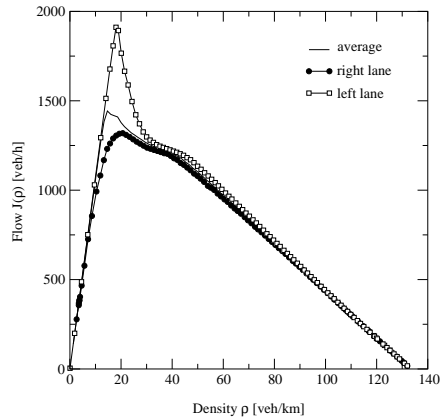,width=0.52\linewidth} } 
\caption{Fundamental diagram of the individual lanes in the inhomogeneous
asymmetric model. Note, that trucks are
not allowed to change the lane.}
\label{asym:inhomo:ban:flow}
\end{center}
\end{figure}

\begin{figure}[h]
\begin{center}
\subfigure{\epsfig{file=\DIR/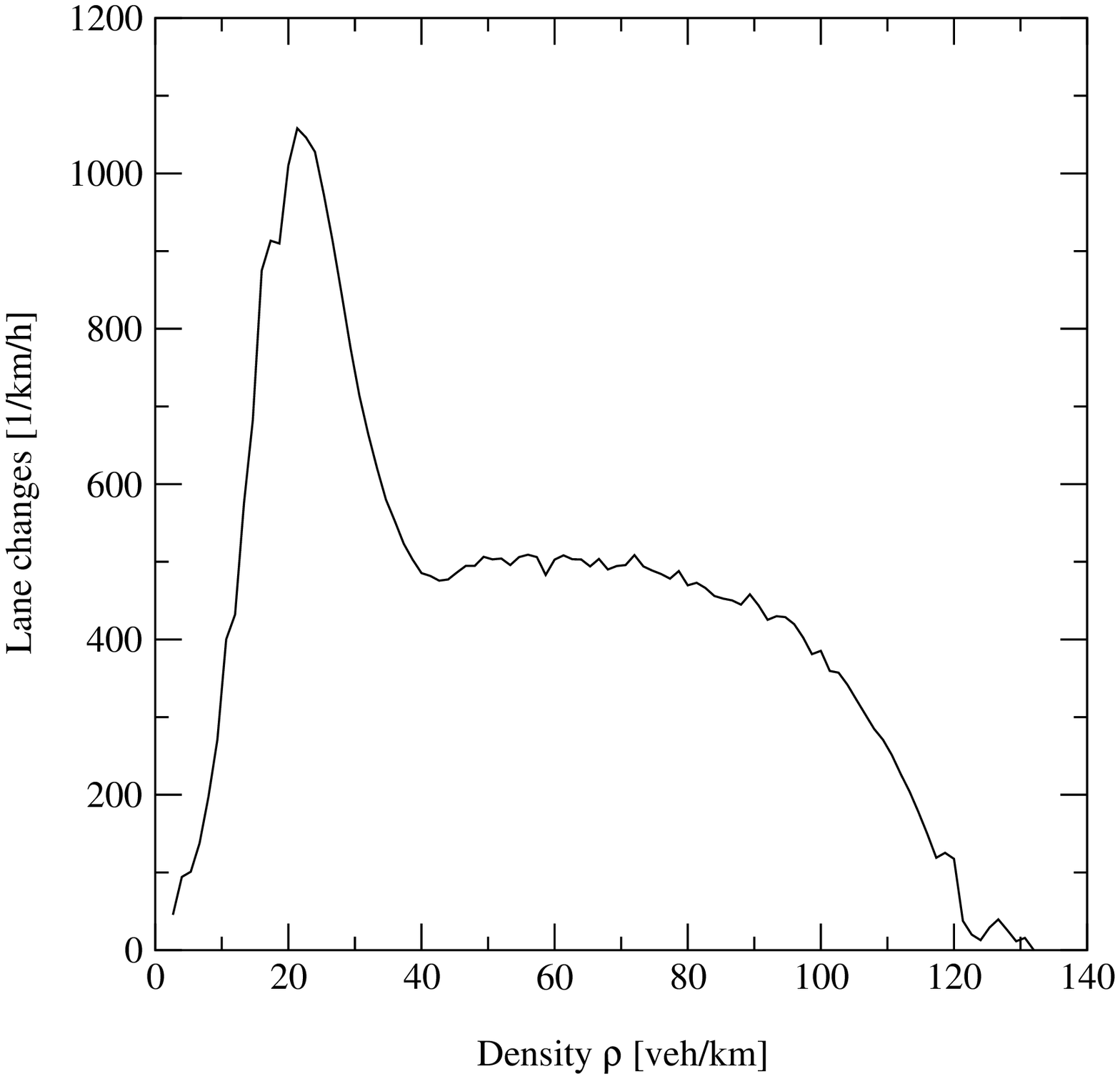,width=0.52\linewidth}
}  
\caption{Lane changes in the inhomogeneous asymmetric model. Note,
that trucks are not allowed to change the lane.}
\label{asym:inhomo:ban:lanechanges}
\end{center}
\end{figure}

\begin{figure}[h]
\begin{center}
\subfigure{\epsfig{file=\DIR/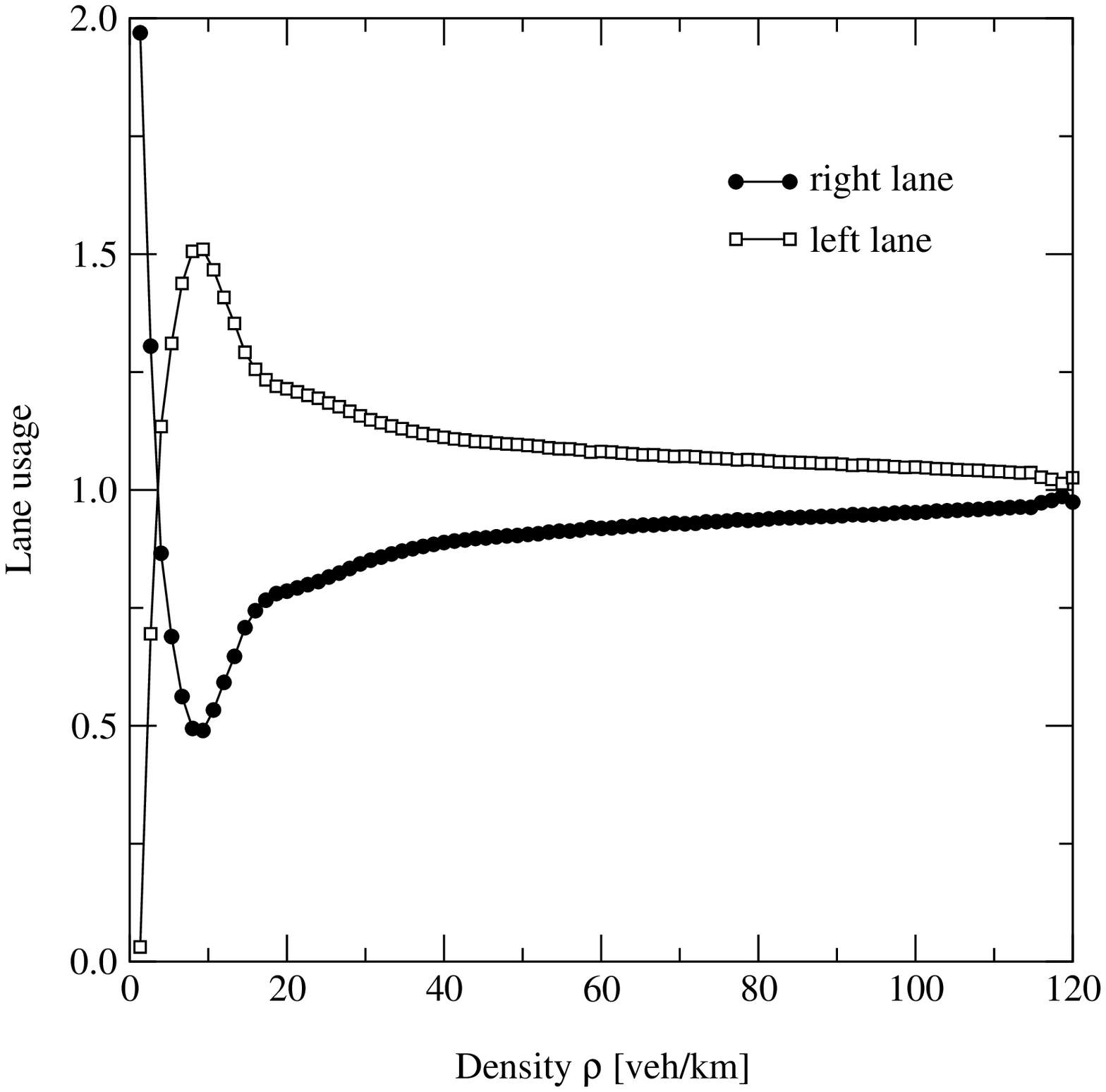,width=0.52\linewidth}
}
\caption{Lane usage in the inhomogeneous asymmetric model. Note, that
trucks are not allowed to change the lane.}
\label{asym:inhomo:ban:laneusage}
\end{center}
\end{figure}

The stability of the plugs is mainly determined by velocity
fluctuations since the distance between two trucks driving side by side
in free flow can only be increased by a repeated deceleration of
one truck in the randomization step of the velocity update.
We therefore distributed the maximum velocity of the cars 
$v_{{\rm max}}$ according to a Gaussian profile with different 
variances and a mean $\bar{v}_{{\rm max}}=20$.
Despite the possibility that now small differences of the maximum
velocity of the trucks are allowed, the plugs remain still very robust.
Even different Gaussian distributions of the maximum velocities of cars
and trucks or special lane changing rules for trucks do not decrease
the stability of the plugs. 
Plugs are therefore no artifact of the discretization but are formed
by the dynamics. Because of the asymmetry of the lane changing
rules, once a truck changed to the left lane, there has to be a large
gap in front on the left and on the right lane to change back. This
increases the time trucks are on the left lane and supports the
formation of plugs.
In contrast to the symmetric model, even one truck on the left lane
is able to dominate the flow of the left lane completely. As long as the truck
stays on the left lane passing is not possible: 
due to the right lane
overtaking ban, like in reality, the fast cars pile up behind the truck
waiting for a possibility to pass. 

In order to avoid the formation of plugs we reintroduce the lane
changing ban for the trucks, so that the trucks now are lane stationary
on the right lane.
Like in the symmetric model, the flow of the right lane is dominated
by trucks only in the vicinity of $\rho_{{\rm max}}$
(Fig.~\ref{asym:inhomo:ban:flow}) which is a consequence of the  
increased robustness of the model regarding slow vehicles.
As a result of the moving hindrances on the right lane, the number of
lane changes is increased compared to the homogeneous model
(Fig.~\ref{asym:inhomo:ban:lanechanges}). Now, lane changes are
necessary even at very small densities to avoid being trapped behind a
truck. Therefore, the lane usage inversion is
increased significantly in the vicinity of $\rho_{{\rm max}}$
(Fig.~\ref{asym:inhomo:ban:laneusage}). 

\begin{figure}[hbt]
\begin{center}
\subfigure{\epsfig{file=\DIR/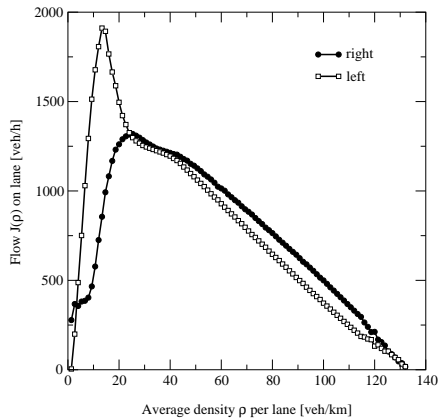,width=0.52\linewidth}
}
\caption{Fundamental diagram of the individual lanes in the asymmetric
model. The density is given as the average density of the individual
lanes. Note, that trucks are not allowed to change the lane.}
\label{asym:inhomo:breakdown}
\end{center}
\end{figure}

Like in the homogeneous model the traffic breakdown is mainly
controlled by the capacity of the left lane
(Fig.~\ref{asym:inhomo:breakdown}). 
At small densities the flow on the left lane is larger than on the right lane
since the density is distributed asymmetrically.
Further increasing the density shifts the excess flow on the right
lane up to its capacity.
At large densities more vehicles are
on the left than on the right lane because of the lane usage inversion,
which results in a larger flow on the right lane.

\section{Local measurements}
\label{section:local}

In the last sections we focused our analysis on two-lane
properties induced by lane changing maneuvers like the number of
lane changes, the flow enhancement and the lane usage inversion.
It turned out that it is possible to produce a realistic lane changing
behavior which can 
reproduce empirical data already by means
of a simple asymmetric rule set. 
Nevertheless, it is an open question whether the vehicle dynamics of the
single-lane model is modified by the introduction of lane
changing rules.

In order to compare the two-lane model with the corresponding
single-lane model, 
we evaluated the simulation data by a virtual inductive loop, i.e., we
measured the speed and the time-headway of the vehicles at a given
link of the lattice. The density is calculated via the relation $\rho
= J/v$ where $J$ and $v$ are the mean flow and the mean velocity of
cars passing the detector in a time interval of $1$ min. In addition
to the aggregated data, the single-vehicle data of each car passing
the detector are also analyzed.

A detailed analysis of the locally measured data of the homogeneous
model shows no difference 
of the two-lane model to the single-lane model on a macroscopic (i.e.,
data averaged over an 
interval of $1$ min) as well as on a microscopic (i.e., single-vehicle
data) level in the congested regime. 
Moreover, there is no difference of the locally measured data between
the two-lanes in the congested regime.
In contrast, in the free flow regime 
the time-headway distribution of the left lane shows an increased
number of small headways compared to the right lane.
The lane usage inversion at small densities increases the car-car
interactions on the left lane which leads to clusters of cars with
small gaps and large velocities.
In the inhomogeneous model the inclusion of disorder
dissolves these clusters so that the difference between the
time-headway distributions to the single-lane model vanishes again.
Moreover, trucks on the right lane reduce the mean velocity in the
free flow regime and shift the
velocity-headway curve to smaller velocities at small distances
compared to the left lane. 
Nevertheless, these results show, that the vehicle dynamics is not
influenced by the lane changing rules but by the traffic state~\footnote{A
detailed analysis of the single-lane model reveals the independence of
the vehicle dynamics in the congested regime even on the maximum velocity.}.

Empirical observations suggest a strong coupling of the lanes in the
synchronized state~\cite{Neubert}. 
In order to quantify the interaction of the lanes, we calculated the
crosscorrelation $cc(\tau)$ of the time-series of measurements of the
density, the 
flow and the velocity on different lanes:

\begin{equation}
cc(x_i,x_j) = \frac{\langle x_i(t)x_j(t + \tau)\rangle - \langle x_i(t)\rangle
  \langle x_j(t + \tau) \rangle}{\sqrt{\langle x_i^{2}(t)\rangle-\langle
    x_i(t)\rangle^{2}}\sqrt{\langle
    x_j^{2}(t+\tau)\rangle-\langle x_j(t+\tau)\rangle^{2}}}. 
\end{equation}

Here $x_i$ denotes either the flow, the density or the velocity in
lane $i$.

\begin{figure}[hbt]
\begin{center}
\subfigure{\epsfig{file=\DIR/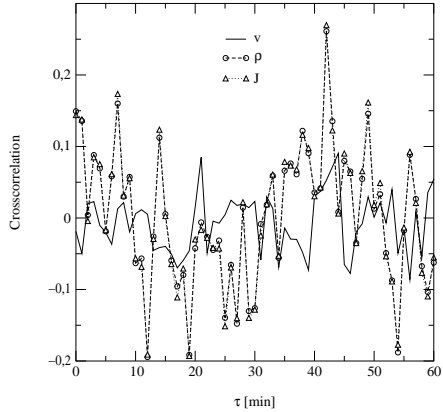,width=0.52\linewidth}
}
\caption{Crosscorrelation of the locally measured flow, density and
velocity of neighboring lanes in the free flow 
regime ($\rho=6~veh/km$) in the asymmetric model.}
\label{asym:crosscorr1}
\end{center}
\end{figure}

\begin{figure}[hbt]
\begin{center}
\subfigure{\epsfig{file=\DIR/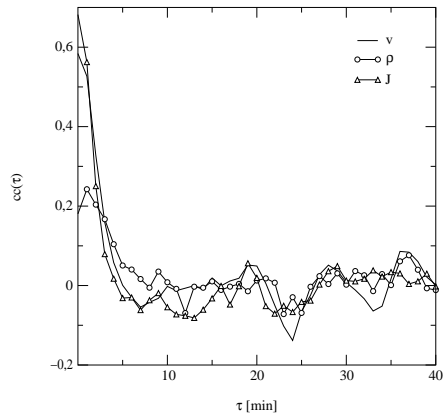,width=0.52\linewidth}
}
\caption{Crosscorrelation of the local measured flow, density and
velocity of neighboring lanes in the
synchronized state ($\rho=50~veh/km$) in the asymmetric model.}
\label{asym:crosscorr2}
\end{center}
\end{figure}

In the free flow state the weak interaction of the lanes results in a
small correlation of all quantities 
(Fig.~\ref{asym:crosscorr1}). Since the flow is mainly controlled by
density fluctuations the density and the flow
measurements show the same correlations. 
In contrast, at larger densities the velocity and the flow on
both lanes is 
strongly correlated ($cc(x_i,x_j) \le 0.6$ at $\tau = 0$), i.e.,
indicating that the system is 
in the synchronized regime. However, due to the large variance of 
the density measurements in the synchronized regime the
density on both lanes is not correlated (Fig.~\ref{asym:crosscorr2}).

\begin{figure}[hbt]
\begin{center}
\subfigure{\epsfig{file=\DIR/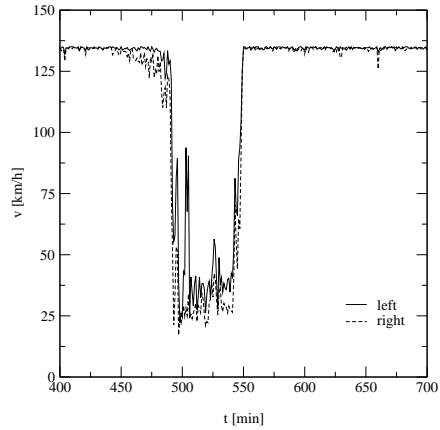,width=0.52\linewidth}
}
\caption{Asymmetric model with open boundary conditions. The large
input rate of $0.35$ of the onramp induces jams on both lanes.}
\label{asym:open_jam}
\end{center}
\end{figure}

\begin{figure}[hbt]
\begin{center}
\subfigure{\epsfig{file=\DIR/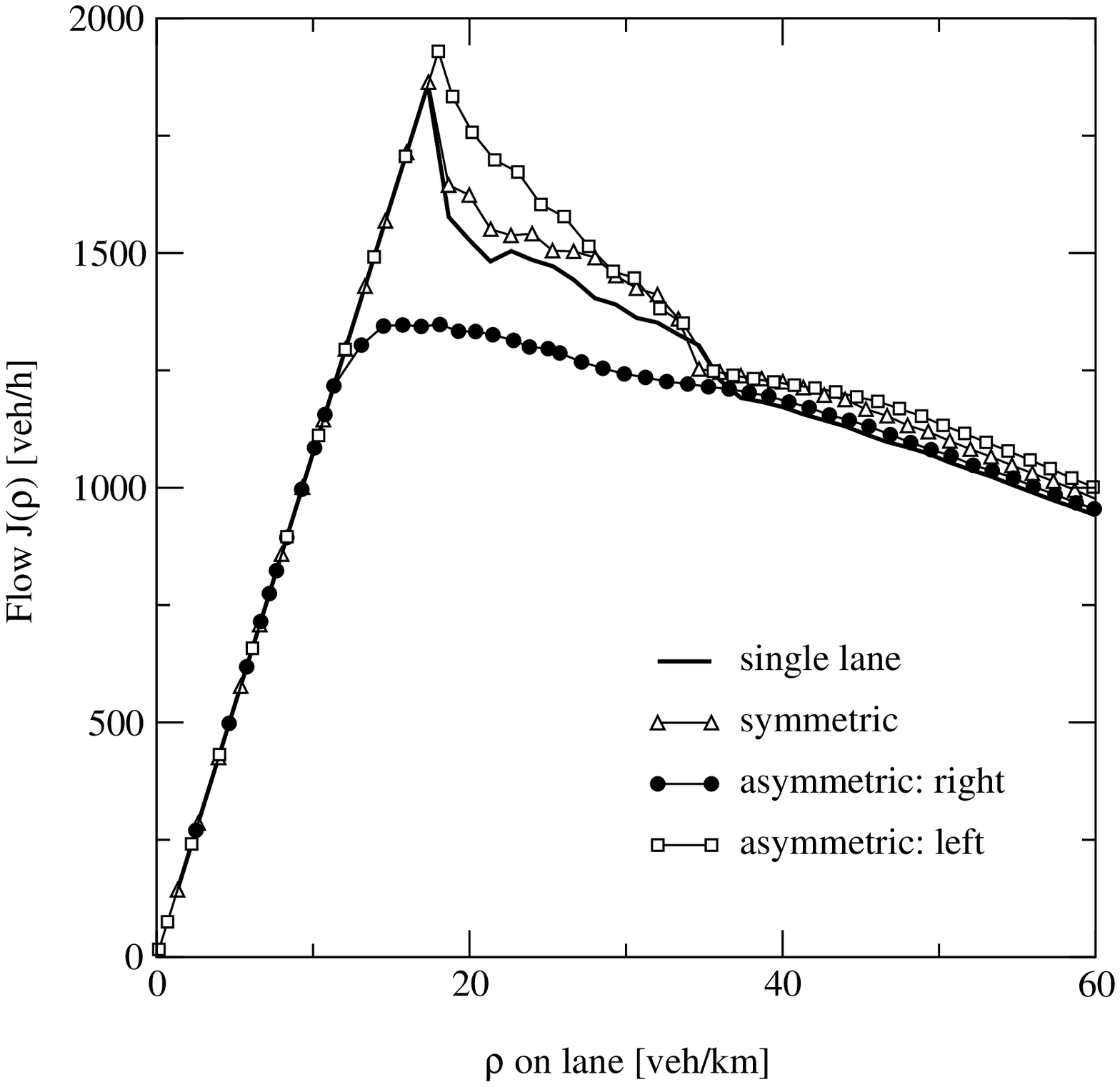,width=0.52\linewidth}
}
\caption{Comparison of the fundamental diagram of the single-lane
model with the homogeneous two-lane models.} 
\label{asym:fundi:compare}
\end{center}
\end{figure}

Lane changes are responsible for the synchronization of the velocities
of different lanes \cite{Kerner2001}. In case of a velocity difference
between the lanes, vehicles are changing to the faster lanes, thus
decreasing the velocity difference. This leads to a strong coupling of
the lanes in the congested regime.  In particular, in the vicinity of
on- and offramps cars enter or leave only the right lane, but both
lanes are disturbed at large in- or outflows.  In order to verify
this, we simulated a system with open boundary conditions and
generated a jam on the right lane by a large input of an onramp.  As a
result, the jam spreads to the left lane, so that the time-series of
the velocity measurements of both, the left and the right lane are
highly correlated. Moreover, the traffic breakdown occurs at the same
time (Fig.~\ref{asym:open_jam}) leading to a synchronization of the
lanes.

\section{Single-lane vs. two-lane traffic}
\label{section:comparison}

A detailed comparison of the single-lane model with the corresponding
two-lane models shows that
it is not possible to increase the flow of a two-lane model to more
than twice
the flow of a single-lane
model by means of an asymmetric lane interaction.
In contrast to symmetric models, the flow of the asymmetric model on
the right lane is decreased whereas the flow on the left lane is
increased compared  
with the single-lane model (Fig.~\ref{asym:fundi:compare}). The right
lane overtaking ban and the right 
lane preference lead to a suboptimal gap usage and, therefore, to a
reduction of the flow on the right lane.
As a result, the mean flow per lane is smaller than the flow of a
homogeneous single-lane model. 

However, the benefits of the two-lane model become visible 
if different types of cars with different
maximum velocities are considered.
(Note, that the trucks are not allowed to change the lane.)
Due to the possibility of overtaking, the system is more robust
regarding disorder.
While the flow of a single-lane model is dominated already by just one 
slow car~\cite{knospe:physicaa,chowdhury:physicaa} (since passing is not
possible), the flow on the right lane of the 
two-lane model is only decreased in the vicinity of $\rho_{{\rm max}}$
compared with the homogeneous model.

\begin{figure}[hbt]
\begin{center}
\subfigure{\epsfig{file=\DIR/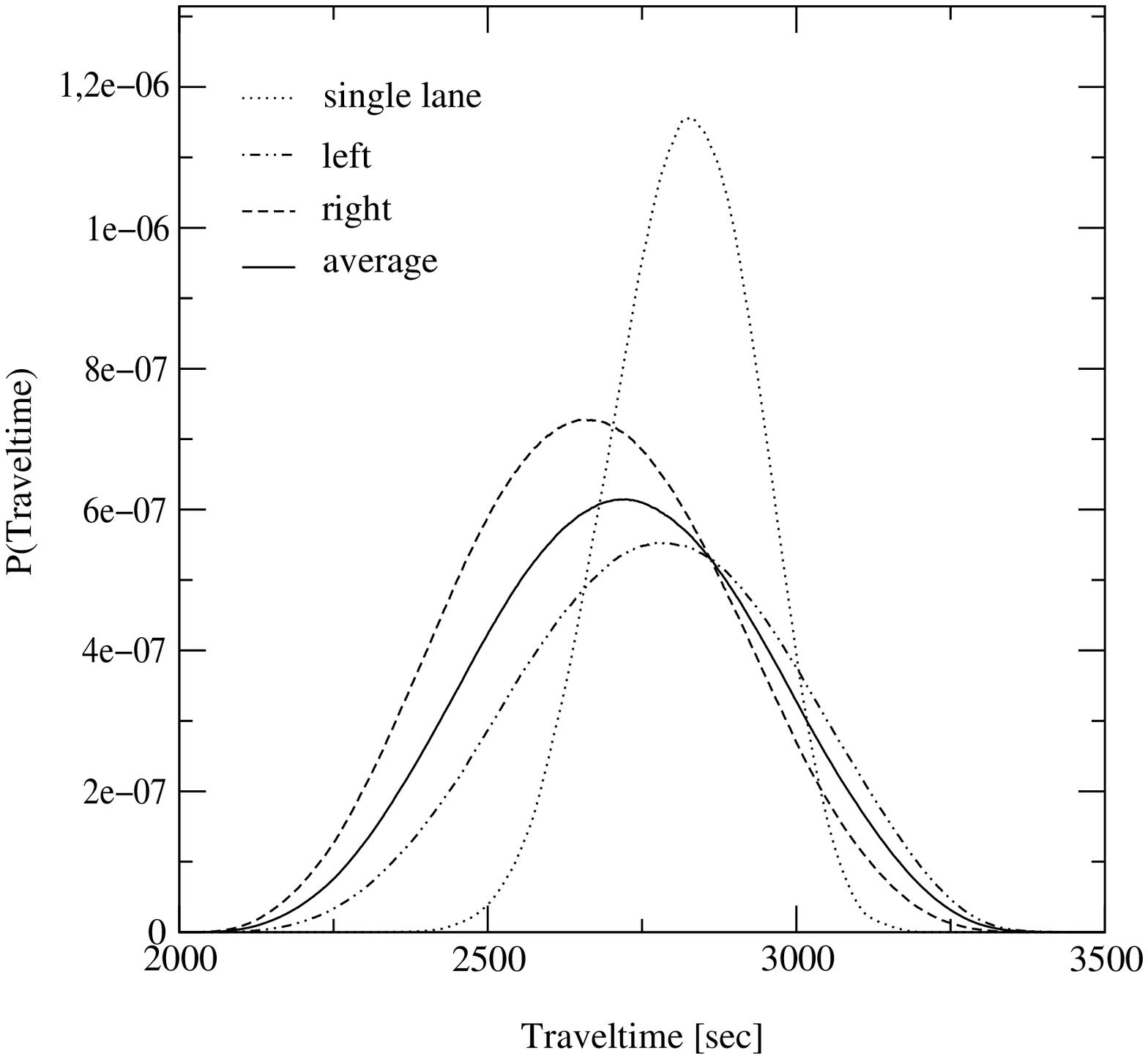,width=0.52\linewidth}
}
\caption{Travel-time distribution of the asymmetric two-lane
model compared with the single-lane model at a density of
$54~veh/km$. The data points were smoothed by a moving
average. The system length is $15$ km.}
\label{asym:traveltime1}
\end{center}
\end{figure}

\begin{figure}[hbt]
\begin{center}
\subfigure{\epsfig{file=\DIR/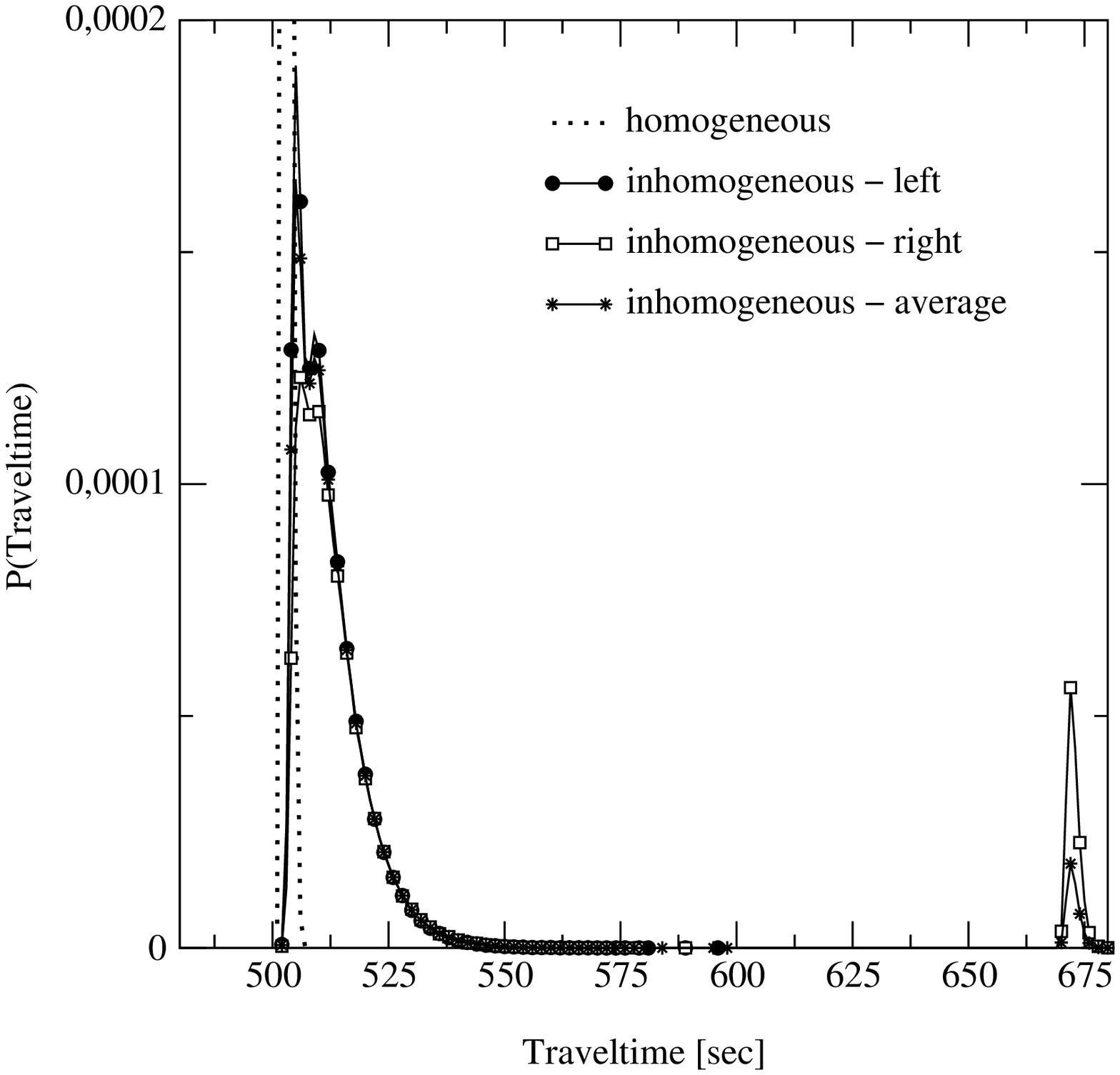,width=0.52\linewidth}
}
\caption{Comparison of the travel-time distributions of the homogeneous 
and the inhomogeneous asymmetric model at a density of $14~veh/km$. The
system length is $15$ km.}
\label{asym:traveltime2}
\end{center}
\end{figure}

Furthermore, the travel-time of a vehicle is reduced considerably by the
lane interaction. The travel-time is measured as the time a vehicle needs 
to pass a given fixed link of the lattice twice. Thus the
travel-time on the right lane is given by vehicles that passed the
measurement section on the right lane. Here it is not distinguished
whether the car started on the left or the right lane and
lane changes during the trip through the system are allowed. 

Compared to the  
single-lane model, the width of the travel-time distribution in the 
congested regime of the two-lane model is larger 
(Fig.~\ref{asym:traveltime1}) but the maximum has moved to smaller
times. On one hand, vehicles which cannot 
change the lane due to an insufficient safety gap lead to larger
travel-times. On the other hand, the possibility to change the lane
reduces the travel-time.
Moreover, the asymmetric distribution of the density leads to
a smaller travel-time on the right lane than on the left lane
Again, the results in the congested regime are independent on the
fraction of trucks since driving with maximum velocity is no longer
possible. 

In the free flow regime, the inclusion of trucks on the right lane is
responsible for a larger mean travel-time compared with the
homogeneous two-lane model (Fig.~\ref{asym:traveltime2}). 
The trucks on the right lane enforce clustering so that the travel-time
of cars trapped behind a truck increases which leads to a second
small peak at large times in the travel-time distribution.
Therefore, the more trucks are in the system, the larger is the variance 
of the travel-time distribution.

\section{Conclusion}
\label{section:conclusion}

We have presented an asymmetric two-lane model with a simple and local
lane changing rule set that shows the
empirically observed lane usage inversion and reproduces the 
density dependence of the number of lane changes.
In order to observe lane usage inversion it is not sufficient to
implement asymmetry into the model by means of disorder, i.e.,
different types of cars. Moreover, the introduction of the right lane
overtaking ban or the right lane preference alone fails to reproduce
the lane usage inversion, too.
Thus, realistic behavior is only obtained by incorporating both a right 
lane preference and a right lane overtaking ban simultaneously into 
the model's rules.

The benefits of the two-lane model compared with the single-lane model
become visible by introducing disorder:
there is no flow 
enhancement of the two-lane model with one type of cars, but the
robustness of the flow regarding slow vehicles can be increased
significantly by means of lane interaction. 
This lane interaction leads to a strong coupling of both lanes in the
synchronized regime, which results in strong correlations of the
time-series of the velocity. 

As a consequence, a jam generated on the right lane by an onramp can branch
to the left lane leading to a congested region on both lanes. 
In order to improve and optimize the throughput of the highway it is
possible to use ramp metering
systems~\cite{papageorgiou1995,salem1995,huberman1999}. 
Here, such systems will help to reduce perturbations of the
right lane and, thus, stabilize the flow even on the left lane.

Finally, a detailed analysis of local measurements shows that the car dynamics
remains unchanged by the introduction of the lane changing rules. 
Therefore, the lane change extension of the single-lane model can be
used for the propagation of traffic state information to all lanes,
e.g., in order to synchronize the lanes, or to trigger distinct traffic 
states (e.g., in the vicinity of on- and offramps).
Moreover, the independence of the velocity update and the lane change
update allows for further improvements of the car dynamics and/or
developments of more sophisticated lane changing rules (e.g., for
bottlenecks).

{\bf Acknowledgments}: 
The authors are grateful to the 
Ministry of Economics and Midsize Businesses, Technology and Transport
for the financial support.
L.~S.~acknowledges support from the Deutsche Forschungsgemeinschaft
under Grant No. SA864/1-2.

\end{document}